\documentstyle[amsmath,amssymb,graphicx]{article}

\def\be{\begin{eqnarray}}
\def\ee{\end{eqnarray}}
\def\nn{\nonumber}
\def\p{\partial}

\def\tr{{\rm tr}\,}
\def\Tr{{\rm Tr}\,}

\hoffset= - 1.0in         

\begin{document}

\baselineskip14pt

\hfill ITEP/TH-08/09

\bigskip

\centerline{\Large{Generation of Matrix Models by $\hat W$-operators
}}

\bigskip

\centerline{A.Morozov and Sh.Shakirov\footnote{morozov@itep.ru;\ shakirov@itep.ru}}

\bigskip

\centerline{{\it ITEP, Moscow, Russia}}

\centerline{{\it MIPT, Dolgoprudny, Russia}}

\bigskip

\centerline{ABSTRACT}

\bigskip

{\footnotesize
We show that partition functions of various matrix models can be obtained
by acting on elementary functions with exponents of $\hat{W}$ operators.
A number of illustrations is given, including the Gaussian Hermitian matrix model,
Hermitian model in external field and the Hurwitz-Kontsevitch model,
for which we suggest an elegant matrix-model representation.
In all these examples, the relevant $\hat{W}$ operators belong
to the $\hat{W}^{(3)}$ algebra.
}

\bigskip

\tableofcontents

\section{Introduction}

These days we witness a renaissance of matrix model theory,
both of its applications and of "theoretical theory".
Matrix models are finally recognized as a source of
new, badly needed, special functions and as a simplified,
still representative, model of entire string/M-theory,
with sophisticated generalized geometries of Calabi-Yau type
behind the vacua structure substituted by a far better studied
geometry of the Riemann surfaces. For summaries of the previous stages
of matrix model theory see, for example, \cite{sumf}-\cite{UFN3}
and references therein. For recent papers with a number of
advanced new developments see \cite{DV}-\cite{sha}.

One of important issues about string theory partition functions
is their generation by canonical procedures from simple canonical objects.
There is a whole variety of such reductionistic properties.
In the case of matrix models one can think of reducing them
to a simpler (more fundamental) {\it theory} -- like that of free fields
on Riemann surfaces and thus to $2d$ conformal and finally to group theory, see
\cite{quiv,UFN3,ammsf,eo} for different stages of this project. One can embed matrix models into group theory
in a somewhat different way, by exploring and exploiting integrability properties of
partition functions \cite{GKM,UFN3} (partition function of a quantum theory is always a kind
of a $\tau$-function \cite{gentau}, and in the case of the eigenvalue
matrix models these are usually $\tau$-functions from the
well studied KP-Toda family, associated with the $\widehat{ U(1) }$ Lie
algebra). One can instead express generic matrix models through
a few simple ones, like the Kontsevich model \cite{GKM},
which has alternative origins in combinatorics
and geometry of moduli spaces, see \cite{ammer}-\cite{sh}.
This is a part of reductionist program within the matrix model
field itself, especially important if one uses it to model the
pertinent features of string theory \cite{ammer}.

In this paper we address another reduction of the same type: from sophisticated to simple $\tau$-functions, but a much
simpler one as compared to meron/instanton decompositions
of \cite{ammer}. Namely, as previously observed in the case of
the Hurwitz-Kontsevich model \cite{hk}, partition functions
can be generated from some trivial $\tau$-functions -- like
$e^{t_1}$ -- by the action of non-trivial generators
$$ \mbox{ Partition Function } = e^{\hat W} \big( \mbox{ Elementary Function } \big) $$
from integrability-preserving $GL(\infty)$ group,
which converts one family of Virasoro-like constraints into another
(it is actually enough to look at the string equations).
See the basics of the underlying theory of equivalent hierarchies in \cite{GLinf}.
Operators $\hat W$ are naturally classified by their spin:
when constructed from free fields, the spin-$k$ operators have the form

$$ {\hat W}^{(k)}(z) \ = \ \sum\limits_{m = -\infty}^{\infty}
{\hat W}^{(k)}_m \ \dfrac{\big(dz\big)^k}{z^{k + m}} \sim \big(\partial\Phi\big)^k(z) $$
\smallskip\\
i.e. are made from the $k$-th powers of the $\widehat{U(1)}$
currents on Riemann surfaces.
In the simplest examples, which we consider in this paper,
the relevant generators are just the next-complicated after
the spin-2 Virasoro ones: the $\hat W^{(3)}$ operators \cite{w3}
(we call them simply $\hat W$ in what follows).
When expressed through the $n \times n$ matrix-valued
background field $\psi$ (the Miwa variable), operators $\hat W^{(k)}$
are differential operators of order $(k - 1)$, so for $k = 3$ they resemble
Laplace operators:

$$ W^{(3)}_{2} = \tr \left( \dfrac{\partial}{\partial \psi} \right)^2 =
\dfrac{\partial^2}{\partial \psi^i_j \partial \psi^j_i} $$
and
$$ W^{(3)}_{0} = \tr \left( \psi \dfrac{\partial}{\partial \psi^T}
\right)^2 - n\ \tr \left( \psi \dfrac{\partial}{\partial \psi^T} \right)
= \psi^j_k \psi^i_l \dfrac{\partial^2}{\partial \psi^i_j \partial \psi^k_l} $$
\smallskip\\
are the simplest illustrations of this property, which plays an important role below. The goal of this paper is just to describe a few examples,
leaving intriguing applications to separate publications.
Some relations to GKM theory \cite{GKM} are immediately obvious,
but even they will be discussed elsewhere --
in order to clearly separate explicit formulas of the
present paper from broader hypotheses and speculations.

\section{Hermitian Matrix Model in external field and operator $\hat W_{2}$}

To begin with, we consider external-field correlators in the Hermitian matrix model

$$C_{k_1, \ldots, k_m} = \int\limits_{N\times N} d\phi\ \ e^{-\Tr \phi^2/2}
\ \Tr (\phi + \psi)^{k_1}\ldots \Tr(\phi + \psi)^{k_m} $$
\smallskip\\
where the external field $\psi$ is a constant Hermitian $N \times N$ matrix and $d \phi = \prod_{i,j} d \phi^{i}_{j}$ is the flat measure. Obviously, correllators $C_{k_1, \ldots, k_m}$ are invariant under conjugation $\psi \mapsto U \psi U^{-1}$, i.e. they are functions of invariant variables $T_k = \Tr \psi^k$.  For example

\[
\begin{array}{l}
C_{1} = T_{1}\\
\\
C_{2} = T_{2}+T_{0}^2\\
\\
C_{3} = T_{3}+3 T_{0} T_{1}\\
\\
C_{1,1} = T_{1}^2+T_{0}\\
\\
C_{2,2} = T_{2}^2+4 T_{2}+2 T_{0}^2 T_{2}+2 T_{0}^2+T_{0}^4\\
\\
C_{1,1,2} = T_{1}^2 T_{2}+T_{0} T_{2}+4 T_{1}^2+T_{0}^2 T_{1}^2+2 T_{0}+T_{0}^3\\
\end{array}
\]
\smallskip\\
Note, that correlators depend on $N$ only through $T_0 = \Tr \psi^0 = N$.
With the help of the shift operator

$$\exp\left( \Tr \phi \dfrac{\partial}{\partial \psi^{T}} \right) f\big(\psi\big) = f\big(\phi + \psi\big) \ \ \ \ \forall f,$$
\smallskip\\
where $\big( \psi^T \big)^i_j = \psi^j_i$ is the transposed matrix,
the integral over $\phi$ can be made Gaussian:
\begin{align*}
\\ C_{k_1, \ldots, k_m} \ \ & = \int d\phi\ \ e^{-\Tr \phi^2/2} \ \Tr (\phi + \psi)^{k_1}\ldots \Tr(\phi + \psi)^{k_m}
\\ & \\ & = \int d\phi\ \ \exp\left(-\Tr \phi^2/2 + \Tr \phi \dfrac{\partial}{\partial \psi^{T}} \right) \ \Tr \psi^{k_1} \ldots \Tr \psi^{k_m}
\\ & \\ & = \exp \left( \dfrac{1}{2} \Tr \left(\dfrac{\partial}{\partial \psi^{T}}\right)^2 \right) \Tr \psi^{k_1} \ldots \Tr \psi^{k_m} = \exp \left( \dfrac{1}{2} \Tr \left(\dfrac{\partial}{\partial \psi}\right)^2 \right) \Tr \psi^{k_1} \ldots \Tr \psi^{k_m} \\ &
\end{align*}
In this way we find an explicit formula for all correlators:

\begin{align}
C_{k_1, \ldots, k_m} = e^{ {\hat W}_{2} / 2} \ T_{k_1} \ldots T_{k_m}
\label{1}
\end{align}
where operator

$$
\hat W_2 = \Tr \left(\dfrac{\partial}{\partial \psi}\right)^2 = \dfrac{\partial^2}{\partial \psi^i_j \partial \psi^j_i }
$$
\smallskip\\
can be called a matrix Laplace operator. It converts invariant (under conjugation of $\psi$) functions into invariant functions, and therefore, can be reduced to the space of such functions, where it acts as a differential operator of second order in invariant variables $T_{k}$. Indeed, by application of the chain rule

$$ \dfrac{\partial}{\partial \psi^i_j} F(T) = \sum\limits_{a = 0}^{\infty} \dfrac{\partial T_{a}}{\partial \psi^i_j} \dfrac{\partial F(T)}{\partial T_a} $$
\smallskip\\
and similarly

$$ \dfrac{\partial^2}{\partial \psi^i_j \partial \psi^j_i } F(T) = \sum\limits_{a,b = 0}^{\infty} \dfrac{\partial T_{a}}{\partial \psi^i_j} \dfrac{\partial T_{b}}{\partial \psi^j_i} \dfrac{\partial^2 F(T)}{\partial T_a \partial T_b} + \sum\limits_{a = 0}^{\infty} \dfrac{\partial^2 T_{a}}{\partial \psi^i_j \partial \psi^j_i} \dfrac{\partial F(T)}{\partial T_a} $$
\smallskip\\
Taking derivatives of traces, it is easy to check that
$$ \dfrac{\partial T_{a}}{\partial \psi^i_j} \dfrac{\partial T_{b}}{\partial \psi^j_i} = ab T_{a + b - 2} $$
and
$$ \dfrac{\partial^2 T_{a}}{\partial \psi^i_j \partial \psi^j_i} = \sum\limits_{k + l = a - 2} (k + l + 2) T_{k} T_{l}  $$
Therefore

\begin{align}
\hat W_2 = \Tr \left(\dfrac{\partial}{\partial \psi}\right)^2 = \sum_{a,b = 0}^{\infty} \left((a+b+2)T_aT_b\frac{\p}{\p T_{a+b+2}} +
ab T_{a+b - 2}\frac{\p^2}{\p T_{a}\p T_{b}}\right)\label{2}
\end{align}
\smallskip\\
Once again, the last identity is true, when the operator acts on invariant functions,
i.e. on functions of time-variables $T_{k}$.
As usual, the partition function is introduced as generating function for all correlators.
It depends on two sets of time-variables, $t_k$ and $T_k$:
\footnote{Through this section, to simplify our formulas and to make closer contact
with \cite{hk} in s.\ref{huk} below,
we omit the factor $1/k$ in Miwa transform $T_k = \Tr \psi^k$,
what, actually, spoils the natural symmetry between $t_k$ and $T_k$.
If $1/k$ is restored, $T_k = \Tr \psi^k / k$, then the exponent in (\ref{3})
acquires its usual form $\exp\big( k t_k T_k \big)$.
In this case, however, one gets an additional factor $\exp\big( t_0 T_0 \big)$. }

$$ Z\big(t|T\big) \ = \ \sum\limits_{m = 0}^{\infty} \dfrac{1}{m!}
\ \sum\limits_{k_1, \ldots, k_m = 0}^{\infty} \ C_{k_1, \ldots, k_m}(T )
\ t_{k_1} \ldots t_{k_m}
= \int d\phi\ \ e^{-\Tr \phi^2/2 + \sum_{k \geq 0} t_{k} \Tr (\phi + \psi)^k } $$
\smallskip\\
It follows from (\ref{1}), that
\begin{equation}
\addtolength{\fboxsep}{5pt}
\boxed{
\begin{gathered}
\ \ \ Z\big(t|T\big) =  e^{ {\hat W}_{2} / 2} e^{\sum_{k \geq 0} t_{k} T_{k}} \ \ \
\end{gathered}
}\label{3}
\end{equation}
\smallskip\\
As one can see, partition function of the Hermitian matrix model
in external field is generated from the trivial function $e^{\Sigma t_{k}T_{k}}$
by the action of generator ${\hat W}_{2}/2$.
Formula (\ref{3}) is quite interesting: an explicit representation for the partition function,
which does not involve matrix integrals.
Also, eqs. (\ref{1}) and (\ref{2}) are very convenient to calculate
particular correlators "by bare hands". For example,

$$ C_{2,2} = \left( 1 + \dfrac{1}{2} \hat W_2 + \dfrac{1}{4} \big(\hat W_2\big)^2 \right) T_{2}^2 $$
\smallskip\\
since all powers of $ \hat W_2 $, higher than 2, annihilate $T_{2}^2$. One finds

$$ \hat W_2 \ T_{2}^2 = 8 T_{2} + 4 T_{0}^2 T_{2}, $$

$$ \big(\hat W_2\big)^2 \ T_{2}^2 = 16 T_{0}^2 + 8 T_{0}^4 $$
and

$$ C_{2,2} = T_{2}^2+4 T_{2}+2 T_{0}^2 T_{2}+2 T_{0}^2+T_{0}^4 $$
\smallskip\\

\section{Gaussian Hermitian Model and operator $\hat W_{-2}$}

\subsection{The main relation, eq.(\ref{Gauss})}

Our next example is the simplest matrix model -- Hermitian matrix model in Gaussian potential:

$$
Z_{G} = \int\limits_{N \times N} d\phi \ e^{ - \Tr \phi^2 / 2 \ + \sum_{k \geq 0} t_k \Tr \phi^k }
$$
\smallskip\\
where the integral is taken over all $N \times N$ Hermitian matrices with flat measure and depends on the set of time-variables $t_k$. Since $\Tr \phi^0 = N$, the dependence on $t_0$ is given by simple multiplicative factor $e^{N t_0}$:

$$
Z_{G} = e^{Nt_0} \int\limits_{N \times N} d\phi \ e^{ - \Tr \phi^2 / 2 \ + \sum_{k > 0} t_k \Tr \phi^k }
$$
\smallskip\\
We are going to demonstrate, that partition function of the Gaussian model is generated by operator ${\hat W}_{-2}$
\begin{equation}
\addtolength{\fboxsep}{5pt}
\boxed{
\begin{gathered}
\ \ \ Z_{G} = e^{ {\hat W}_{-2} / 2 } e^{N t_0} \ \ \
\end{gathered}
}\label{Gauss}
\end{equation}
where
\begin{align}
{\hat W}_{-2} = \sum\limits_{a,b = 0}^{\infty} \left( a b t_{a} t_{b} \dfrac{\partial}{\partial t_{a + b - 2}} + (a + b + 2) t_{a + b + 2} \dfrac{\partial}{\partial t_{a}} \dfrac{\partial}{\partial t_{b}} \right)
\label{W-2}
\end{align}
\smallskip\\
Note, that ${\hat W}_{2}$ in (\ref{3}) is acting on $T$, while ${\hat W}_{-2}$ in (\ref{Gauss}) -- on $t$-variables. To prove (\ref{Gauss}), we make a Miwa transform -- introduce an $n \times n$ Hermitian matrix $\psi$, such that

$$t_k = \dfrac{1}{k} \tr \psi^{-k}, \ \ \ k > 0$$
\smallskip\\
It is important that the size $n$ of matrix $\psi$ is absolutely independent of the initial size $N$ of matrix $\phi$, because interaction terms in the action do not involve addition or multiplication of these matrices (this was not the case in s.2). This is the usual feature of Kontsevich-like matrix models \cite{GKM}, what emphasizes relation between the subject of this paper and GKM theory. Consequently, we distinguish operations $\tr$ and $\Tr$, which denote traces of matrices of sizes $n$ and $N$, respectively.

After the transform, the left hand side of (\ref{Gauss}) takes form

$$
Z_{G} = e^{Nt_0} \int\limits_{N \times N} d\phi \ e^{ - \Tr \phi^2 / 2 } \ \exp \left( \sum\limits_{k = 1}^{\infty} \dfrac{1}{k} \tr \psi^{-k} \Tr \phi^k \right)
$$
\smallskip\\
Note, that $t_0$ is not affected by this transform and remains a free parameter. Using the identity
$$\det \big( I \otimes I - \psi^{-1} \otimes \phi \big) = \exp \Big( \mbox{tr} \log \big( I \otimes I - \psi^{-1} \otimes \phi \big) \Big) = \exp \left( - \sum\limits_{k = 1}^{\infty} \dfrac{1}{k} \tr \psi^{-k} \Tr \phi^k \right) $$
\smallskip\\
the interaction terms are written in determinantal form:

\begin{align}
Z_{G} =  e^{Nt_0} \int\limits_{N \times N} d\phi \ e^{ - \Tr \phi^2 / 2 } \ \dfrac{1}{\det \big( I \otimes I - \psi^{-1} \otimes \phi \big)} = e^{Nt_0} \int\limits_{N \times N} d\phi \ e^{ - \Tr \phi^2 / 2 } \ \dfrac{\big( \det \psi \big)^N }{\det \big( \psi \otimes I - I \otimes \phi \big)}
\label{LeftHandSide}
\end{align}
\smallskip\\
where $(\phi \otimes \psi)^{i \mu}_{j \nu } = \phi^{i}_{j} \psi^{\mu}_{\nu} $ is the tensor product of two matrices, regarded as a $nN \times nN$ matrix.  Note, that $i, j$ and $\mu, \nu$ are different sorts of indices: $i$ and $j$ take values $1, \ldots, N$, while $\mu$ and $\nu$ take values $1, \ldots, n$. Eq. (\ref{LeftHandSide}) is the expression that we need for the left hand side of (\ref{Gauss}).

Now let us calculate the right hand side. Operator ${\hat W}_{-2}$ can be expressed as a differential operator of the second order in terms of $\psi$ and $t_0$. However, we can always substitute the derivatives with respect to $t_0$ by factors of $N$. A straightforward application of the chain rule, similarly to the previous sections, gives

$$
{\hat W}_{-2} = \tr \left( \dfrac{\partial^2}{\partial \psi^2} - \dfrac{N}{\psi} \right) = \big( \det \psi \big)^{N} \tr \left( \dfrac{\partial}{\partial \psi} \right)^2 \big( \det \psi \big)^{-N}
$$
\smallskip\\
Its exponential is easy to find:

$$
\exp \left( \dfrac{1}{2} {\hat W}_{-2} \right) = \big( \det \psi \big)^{N} \exp \left( \dfrac{1}{2} \tr \left( \dfrac{\partial}{\partial \psi} \right)^2 \right) \big( \det \psi \big)^{-N}
$$
\smallskip\\
Using the identity

$$
\exp \left( \dfrac{1}{2} \tr \left( \dfrac{\partial}{\partial \psi} \right)^2 \right) = \int\limits_{n \times n} d \phi \ \exp \left( -\dfrac{1}{2} \tr \phi^2 + \tr \left( \phi \dfrac{\partial}{\partial \psi} \right) \right)
$$
\smallskip\\
and the properties of the shift operator, we obtain:

\begin{align*}
\exp \left( \dfrac{1}{2} {\hat W}_{-2} \right) e^{N t_0} \ & = \ \int\limits_{n \times n} d \phi \ \exp \left( -\dfrac{1}{2} \tr \phi^2 + \tr \left( \phi \dfrac{\partial}{\partial \psi} \right) \right) \dfrac{e^{N t_0}}{\big( \det \psi \big)^N} \\ & \\ & = \ e^{N t_0} \int\limits_{n \times n} d \phi \ e^{- \tr \phi^2/2} \left( \dfrac{\det \psi}{\det \big( \psi + \phi \big)} \right)^N
\end{align*}
\smallskip\\
This is what we get for the right hand side of (\ref{Gauss}). At first sight, this seems different from (\ref{LeftHandSide}):

$$ \int\limits_{N \times N} d\phi \ e^{ - \Tr \phi^2 / 2 } \ \dfrac{\big( \det \psi \big)^N }{\det \big( \psi \otimes I - I \otimes \phi \big)} \ \ \ \mbox{vs.} \ \ \ \int\limits_{n \times n} d \phi \ e^{- \tr \phi^2/2} \left( \dfrac{\det \psi}{\det \big( \psi + \phi \big)} \right)^N $$
\smallskip\\
even the integration goes over matrices of different size.
However, in fact these two integrals are equal
and we prove this fact in the following subsection.

\pagebreak

\subsection{Eq.(\ref{Gauss}) from Faddeev-Popov trick}
We now prove (\ref{Gauss}) by proving the identity between these integrals:

$$ \int\limits_{N \times N} d\phi \ e^{ - \Tr \phi^2 / 2 } \ \dfrac{\big( \det \psi \big)^N }{\det \big( \psi \otimes I - I \otimes \phi \big)} \ \ \ = \ \ \ \int\limits_{n \times n} d \phi \ e^{- \tr \phi^2/2} \left( \dfrac{\det \psi}{\det \big( \psi + \phi \big)} \right)^N $$
\smallskip\\
or in a more symmetric form

\begin{equation}
\addtolength{\fboxsep}{5pt}
\boxed{
\begin{gathered}
\ \ \ \int\limits_{N \times N} d \phi \ e^{ - \Tr \phi^2 / 2} \dfrac{1}{\det\big(  \psi \otimes I - I \otimes \phi \big)} = \int\limits_{n \times n} d \phi \ e^{ - \tr \phi^2 / 2} \dfrac{1}{\det\big( \psi \otimes I + \phi \otimes I \big)} \ \ \
\end{gathered}
}\label{Identity}
\end{equation}
\smallskip\\
We will use Faddeev-Popov's trick, i.e. representation of $\det^{-1}$ as a Gaussian integral
over auxiliary fields:

$$ \int\limits_{N \times N} d \phi \ e^{- \Tr \phi^2 / 2}
\dfrac{1}{\det\big( \psi \otimes I - I \otimes \phi \big)}
= \int\limits_{N \times N} d \phi \int d b d c \
\exp \Big( - \dfrac{1}{2} \Tr \phi^2 + b_{\mu i}
\big( \psi^{\mu}_{\nu} \delta^{i}_{j} - \delta^{\mu}_{\nu} \phi^{i}_{j} \big) c^{\nu j} \Big) $$

$$ \int\limits_{n \times n} d \phi \ e^{- \tr \phi^2 / 2}
\dfrac{1}{\det\big( I \otimes \psi + I \otimes \phi \big)}
= \int\limits_{n \times n} d \phi \int d b d c \
\exp \Big( - \dfrac{1}{2} \tr \phi^2 + b_{\mu i}
\big( \psi^{\mu}_{\nu} \delta^{i}_{j} + \phi^{\mu}_{\nu} \delta^{i}_{j} \big) c^{\nu j} \Big) $$
\smallskip\\
Here $b_{\mu i}$ and $c^{\nu j} = b^{*}_{\nu j}$ are bosonic (since determinant stands in denominator)
Faddeev-Popov fields. Integrals over $\phi$ are Gaussian. After $\phi$ is integrated out, we obtain

$$ \int\limits_{N \times N} d \phi \ e^{- \Tr \phi^2 / 2}
\dfrac{1}{\det\big( \psi \otimes I - I \otimes \phi \big)}
= \int d b d c \ \exp \Big( \dfrac{1}{2} \ b_{\mu i} b_{\nu j} c^{\mu j} c^{\nu i}
+ b_{\mu j} \psi^{\mu}_{\nu} c^{\nu j} \Big) $$

$$ \int\limits_{n \times n} d \phi \ e^{- \tr \phi^2 / 2}
\dfrac{1}{\det\big( I \otimes \psi + I \otimes \phi \big)} = \int d b d c \
\exp \Big( \dfrac{1}{2} \ b_{\mu i} b_{\nu j} c^{\mu j} c^{\nu i} + b_{\mu j}
\psi^{\mu}_{\nu} c^{\nu j} \Big) $$
\smallskip\\
i.e. the integrals become the same. Thus, our identity

$$ \int\limits_{N \times N} d \phi \ e^{ - \Tr \phi^2 / 2}
\dfrac{1}{\det\big(  \psi \otimes I - I \otimes \phi \big)}
= \int\limits_{n \times n} d \phi \ e^{ - \tr \phi^2 / 2}
\dfrac{1}{\det\big( \psi \otimes I + \phi \otimes I \big)} $$
\smallskip\\
is true. Therefore, representation (\ref{Gauss}) is valid.

At that point it is worth mentioning, that there are other identities between Gaussian integrals,
similar to (\ref{Identity}).
For example, there is a direct analogue of (\ref{Identity}),
with determinants standing in the numerator:

$$ \int\limits_{N \times N} d \phi \ e^{ - \Tr \phi^2 / 2} \
\det\big(  \psi \otimes I - I \otimes \phi \big) = \int\limits_{n \times n} d \phi \
e^{ + \tr \phi^2 / 2} \ \det\big( \psi \otimes I + \phi \otimes I \big)
= \int\limits_{n \times n} d \phi \ e^{ + \tr \phi^2 / 2} \ \det\big( \phi + \psi \big)^N $$
\smallskip\\
This identity is well-known as equivalence \cite{Che}
between the Gaussian model and logarithmic Kontsevich model
and is usually proved by orthogonal polynomial techniques \cite{UFN3}.
Faddeev-Popov's trick seems to be a more economic and elegant way to prove such identities.
When determinant stands in the numerator, one only needs to consider \emph{grassmanian}
Faddeev-Popov ghosts $b_{\mu i}$ and $c_{\nu j}$, as it is usually done in Yang-Mills theory.
Remarkably, identity breaks down, if determinants are raised to any other power,
different from $\pm 1$. \linebreak

\subsection{Eq.(\ref{Gauss}) from Virasoro constraints}

There are different other ways to derive (\ref{Gauss}). For example, Gaussian partition function satisfies a consistent system of linear differential equations called Virasoro constraints

$$ \dfrac{\partial}{\partial t_{b}} Z_G = {\hat L}_{b - 2} Z_G = \left( \sum\limits_{a = 0}^{\infty} a t_{a} \dfrac{\partial}{\partial t_{a + b - 2}} + \sum\limits_{i + j = b - 2} \dfrac{\partial}{\partial t_{i}} \dfrac{\partial}{\partial t_{j}} \right) Z_G, \ \ \ b \geq 1; \ \ \  \dfrac{\partial}{\partial t_0} Z_G = N Z_G $$
\smallskip\\
and (\ref{Gauss}) is their direct corollary. Indeed, summing by $b$ from 1 to infinity with weight $b t_b$, we obtain
\begin{align}
 {\hat D} Z_G = {\hat W}_{-2} Z_G
\label{DW}
\end{align}
Two operators appear in this equality: the degree (dilatation) operator
\begin{align} {\hat D} = \sum\limits_{q = 0}^{\infty} q t_{q} \dfrac{\partial}{\partial t_{q}} = {\hat L}_0 - N^2 \label{Dilatation} \end{align}
and our familiar $W_{-2}$-operator

$$ {\hat W}_{-2} = \sum\limits_{a,b = 0}^{\infty} \left( a b t_{a} t_{b} \dfrac{\partial}{\partial t_{a + b - 2}} + (a + b + 2) t_{a + b + 2} \dfrac{\partial}{\partial t_{a}} \dfrac{\partial}{\partial t_{b}} \right) $$
\smallskip\\
with commutation relation
\begin{align}
{\hat D} {\hat W}_{-2} - {\hat W}_{-2} {\hat D} = 2 {\hat W}_{-2}
\label{DWComm}
\end{align}
Notice, that $Z_G$ is graded by the total $t$-degree:

$$ Z_G = \sum\limits_{s = 0}^{\infty} Z_G^{(s)}, \ \ \ \ Z_G^{(s)} = \sum\limits_{m = 0}^{\infty} \ \sum\limits_{k_1 + \ldots + k_m = s} \left< \tr \phi^{k_1} \ldots \tr \phi^{k_m} \right> \ \dfrac{t_{k_1} \ldots t_{k_m}}{m!} $$
\smallskip\\
Operator ${\hat D}$ preserves this grading:
$$ {\hat D} Z_G^{(s)} = s Z_G^{(s)} $$
\smallskip\\
As follows from (\ref{DW}), operator ${\hat W}_{-2}$ respects the grading in the following sense:

$$
 {\hat W}_{-2} Z_G^{(s)} = (s + 2) Z_G^{(s + 2)}
$$
\smallskip\\
and this implies that graded components of $Z_G$ are generated, one by one, from the lowest component:

$$ Z_G^{(2)} = \dfrac{1}{2} {\hat W}_{-2} Z_G^{(0)}$$

$$ Z_G^{(4)} = \dfrac{1}{2 \cdot 4} \left( {\hat W}_{-2} \right)^2 Z_G^{(0)} $$

$$ \ldots \ldots \ldots $$

$$ Z_G \ = \ Z_G^{(0)} + \dfrac{1}{2} {\hat W}_{-2} Z_G^{(0)} + \dfrac{1}{2 \cdot 4} \left( {\hat W}_{-2} \right)^2 Z_G^{(0)} + \dfrac{1}{2 \cdot 4 \cdot 6} \left( {\hat W}_{-2} \right)^3 Z_G^{(0)} + \ldots \ = \ e^{{\hat W}_{-2} / 2} Z_G^{(0)} $$
\smallskip\\
where $ Z_G^{(0)} $ obviously equals $ e^{N t_{0}} $, so that we derived (\ref{Gauss}) once again.

\subsection{Non-Gaussian models and operators $\hat W_{-p}$}

It is tempting to generalize the above Virasoro derivation of (\ref{Gauss}) to non-Gaussian partition functions

$$
Z_{NG} = \int\limits_{N \times N} d\phi \ e^{ - \Tr \phi^p / p \ + \sum_{k \geq 0} t_k \Tr \phi^k }, \ \ \ p \geq 2
$$
\smallskip\\
since Virasoro constraints for these models are not very complicated:
$$ \dfrac{\partial}{\partial t_{b}} Z_{NG}
= \left( \sum\limits_{a = 0}^{\infty} a t_{a} \dfrac{\partial}{\partial t_{a + b - p}}
+ \sum\limits_{i + j = b - p} \dfrac{\partial}{\partial t_{i}}
\dfrac{\partial}{\partial t_{j}} \right) Z_{NG}, \ \ \ b \geq (p - 1) $$
However, in the non-Gaussian case Virasoro constraints are labeled by
$b \geq (p - 1)$, not by $b \geq 1$. Immediate consequence of this is that $Z_{NG}$
is no longer fixed unambiguously  by Virasoro constraints alone:
some additional requirements should be imposed
(see \cite{DV} for the best studied Dijkgraaf-Vafa example).
The technical procedure from s.3.3 is also inapplicable, because
it is impossible to sum by $b$ from 1 to infinity. Instead, one can sum from $(p - 1)$
to infinity and obtain
\begin{align}
 {\hat D}_{+} Z_G = {\hat W}_{-p} Z_G
\label{DpW}
\end{align}
where
$$ {\hat D}_{+} = \sum\limits_{b = p - 1}^{\infty} b t_{b} \dfrac{\partial}{\partial t_{b}} = {\hat D} - \sum\limits_{b = 0}^{p - 2} b t_{b} \dfrac{\partial}{\partial t_{b}}$$
and
$$ {\hat W}_{-p} = \sum\limits_{a,b = 0}^{\infty} \left( a b t_{a} t_{b} \dfrac{\partial}{\partial t_{a + b - p}} + (a + b + p) t_{a + b + p} \dfrac{\partial}{\partial t_{a}} \dfrac{\partial}{\partial t_{b}} \right) $$
\smallskip\\
Operator ${\hat D}$ is the degree operator (\ref{Dilatation}) in all $t$-variables and satisfies
\begin{align}
{\hat D} {\hat W}_{-p} - {\hat W}_{-p} {\hat D} = p {\hat W}_{-p}
\label{DWNComm}
\end{align}
while ${\hat D}_{+}$ does not satisfy (\ref{DWNComm}) and has the meaning of degree operator in variables $t_{i}$ with $i \geq (p - 2)$ -- only a part of all variables. Such a partial grading is not very useful, since ${\hat W}_{-p}$ does not respect this grading. For these reasons, we can not deal with (\ref{DpW}) as we did with (\ref{DW}). Some additional ideas are required to obtain a $e^W$ representation for non-Gaussian, in particular, the Dijkgraaf-Vafa partition functions.

\section{Hurwitz-Kontsevich Model and operator $\hat W_{0}$ \label{huk}}

In the previous sections, we converted a matrix integral into exponent
of a ${\hat W}$-operator, acting on a simple function.
In this section, an inverse problem is considered: namely,
conversion of the $e^{\hat W}$ formula into a matrix integral.
We discuss this topic with the example of Hurwitz-Kontsevich
partition function \cite{hk}, since historically the $e^{\hat W}$-representation
for this function was found \emph{before} the matrix integral.
In result, we obtain an interesting matrix model representation for $Z_{HK}$.

According to \cite{hk}, $Z_{HK}$ depends on the time-variables $p_k$
and additional deformation parameter $t$ via
\begin{align}
\addtolength{\fboxsep}{5pt}
\boxed{
\begin{gathered}
Z_{HK}(p) = e^{ t {\hat W}_{0} / 2} e^{ p_1 }
\end{gathered}
}\label{4}
\end{align}
where
\begin{align}
{\hat W}_0 = \sum_{a,b = 1}^{\infty} \left((a+b)p_ap_b\frac{\p}{\p p_{a+b}} + abp_{a+b}\frac{\p^2}{\p p_a\p p_b}\right)
\label{5}
\end{align}
Existence of such a formula suggests to look for a matrix integral, responsible for the appearance of ${\hat W}_{0}$. We find this integral in several steps: first, we find an approximate matrix integral, then calculate a few corrections and finally conjecture an exact matrix integral. In the spirit of the previous example, we introduce a Miwa variable -- an $n \times n$ matrix $\psi$, such that
$$ p_k = \tr \psi^k $$
Note, that conventionally Miwa transform is defined as $t_k = \tr \psi^k / k$, but we use the rescaled times $p_k = k t_k$ as in \cite{hk}. Operator ${\hat W}_0$ can be expressed as a differential operator of second order in terms of $\psi$. Using the chain rule, just like in section 2, we obtain:
$$ \dfrac{\partial^2}{\partial \psi_{ij} \partial \psi_{kl}} F(p) = \sum\limits_{a,b = 1}^{\infty} \dfrac{\partial p_{a}}{\partial \psi_{ij}} \dfrac{\partial p_{b}}{\partial \psi_{kl}} \dfrac{\partial^2 F(p)}{\partial p_a \partial p_b} + \sum\limits_{a = 1}^{\infty} \dfrac{\partial^2 p_{a}}{\partial \psi_{ij} \partial \psi_{kl}} \dfrac{\partial F(p)}{\partial p_a} $$
and operator (\ref{5}) is reproduced if we contract this relation with $\psi_{kj} \psi_{il}$:
\begin{align}
{\hat W}_0 = \psi_{kj} \psi_{il} \dfrac{\partial^2}{\partial \psi_{ij} \partial \psi_{kl}} = \tr \left( \psi \dfrac{\partial}{\partial \psi^{T}} \right)^2 - n \tr \left( \psi \dfrac{\partial}{\partial \psi^{T}} \right)
\label{6}
\end{align}
Let us remind, as usual, that identity
$$
\psi_{kj} \psi_{il} \dfrac{\partial^2}{\partial \psi_{ij} \partial \psi_{kl}} = \sum_{a,b = 1}^{\infty} \left((a+b)p_ap_b\frac{\p}{\p p_{a+b}} + abp_{a+b}\frac{\p^2}{\p p_a\p p_b}\right)
$$
is true, when the operator acts on invariant functions, i.e. on functions of time-variables $p_{k}$.

\subsection{Approximate matrix integral}

Having a differential operator in terms of $\psi$, one can rewrite its exponent as a matrix integral over auxiliary matrix $\phi$. Unfortunately, it is not possible to literally apply the method of the previous section -- the identity one would use for this purpose is not quite true:

\begin{align}
\int d \phi \ \exp\left( - \dfrac{1}{2t} \tr \phi^2 + \tr \phi\,\psi \dfrac{\partial}{\partial \psi^{T}} \right) \ \neq \ \exp\left( \dfrac{t}{2} \ \tr \left( \psi \dfrac{\partial}{\partial \psi^{T}} \right)^2 \right)
\label{BadIdent}
\end{align}
\smallskip\\
This is because operator ${\hat A} = \psi \dfrac{\partial}{\partial \psi^{T}}$ is more complicated, than the previously considered operator $\dfrac{\partial}{\partial \psi^{T}}$. Its components do not commute, forming a non-abelian $GL(n)$ algebra

\begin{align}
 {\hat A}^i_j {\hat A}^k_l - {\hat A}^k_l {\hat A}^i_j = \delta^k_j {\hat A}^i_l - \delta^i_l {\hat A}^k_j
\label{GLn}
\end{align}
\smallskip\\
As a result of this, (\ref{BadIdent}) breaks down already in the 2nd order of perturbation theory in $t$:

\be
 \int d\phi \exp\left( - \dfrac{1}{2t} \tr \phi^2 \right) \left(\tr \phi {\hat A}\right)^4
&=&\left[ \hat A^i_j \hat A^j_i \hat A^k_l \hat A^l_k
+ \hat A^i_j \hat A^k_l \hat A^j_i \hat A^l_k
+ \hat A^i_j \hat A^k_l \hat A^l_k \hat A^j_i \right] t^2 = \nn \\ & \nn \\ & = & \left[ 3\left(\tr\hat A^2\right)^2 + (\tr\hat A)^2 - N \tr\hat A^2 \right] t^2
\label{A4dev}
\ee
\smallskip\\
i.e. two additional terms appear at the r.h.s. The leading contribution $3\left(\tr\hat A^2\right)^2$ is what one would expect, if identity (\ref{BadIdent}) was true. The other two terms are due to non-abelian nature of operators ${\hat A}^i_j$ and can be considered as subleading contributions. Following this line of thinking, we state that (\ref{BadIdent}) does not hold exactly, but holds \emph{approximately} and can be considered as a 0-th approximation to a correct identity:

\begin{align}
\int d \phi \ \exp\left( - \dfrac{1}{2t} \tr \phi^2 + \mbox{ corrections } + \tr \phi {\hat A} \right) \ = \ \exp\left( \dfrac{t}{2} \ \tr {\hat A}^2 \right)
\label{GoodIdent}
\end{align}
\smallskip\\
In the next section we will show, that corrections are suppressed by powers of $t$. Having this in mind, we use (\ref{GoodIdent}) to obtain an (approximate) Kontsevich-Hurwitz matrix integral:

$$ e^{t \hat W_0 /2 } e^{\tr \psi} = \exp\left( \dfrac{t}{2} \tr {\hat A}^2 - \dfrac{nt}{2} \tr {\hat A} \right) e^{\tr \psi} \approx \int\limits_{n \times n} d \phi \ \exp\left( - \dfrac{1}{2t} \tr \phi^2 + \tr ( \phi {\hat A} ) - \dfrac{nt}{2} \tr {\hat A} \right) e^{\tr \psi} $$
\smallskip\\
where the integral is taken over all $n \times n$ Hermitian matrices with conventional measure. Using

$$ \exp \left( \tr M {\hat A} \right) f\big(\psi\big) =  \exp \left( \tr M \psi \dfrac{\partial}{\partial \psi^{T}} \right) f\big(\psi\big) =  f\big(e^M \psi\big) \ \ \ \forall M, f, $$
\smallskip\\
we finally obtain
\begin{equation}
\addtolength{\fboxsep}{5pt}
\boxed{
\begin{gathered}
\ \ \ Z_{HK} = e^{t \hat W_0 /2 } e^{\tr \psi} \approx \int\limits_{n \times n} d \phi \ \exp\left( - \dfrac{1}{2t} \tr \phi^2 + e^{-nt/2} \tr \big( e^{\phi} \psi \big) \right) \ \ \
\end{gathered}
}\label{ApproxHurwitz}
\end{equation}
\smallskip\\
Emerging matrix model (\ref{ApproxHurwitz}) with exponential potential in the presence of background field $\psi$, despite being only approximate, is very interesting. Its equation of motion

$$ \dfrac{\partial}{\partial \phi^i_j} \left( - \dfrac{1}{2t} \tr \phi^2  + e^{-nt/2} \tr \big( e^{\phi} \psi \big) \right) = - \dfrac{1}{t} \phi^j_i + e^{-nt/2} \left( \psi  e^{\phi} \right)^j_i = 0 $$
\smallskip\\
or briefly $ \phi e^{-\phi} = t e^{-Nt/2} \psi $ is a transcendental equation on matrix $\phi$, solved by Lambert function:
$$ \phi_{\mbox{cl}} \big( \psi \big) = y + y^2 + \dfrac{3}{2} y^3 + \dfrac{8}{3} y^4 + \ldots = \sum\limits_{m = 1}^{\infty} \dfrac{m^{m-1}}{m!} y^m, \ \ \ \ y = t e^{-Nt/2} \psi $$
which -- naturally -- plays a big role in the still underdeveloped and mysterious theory of the Hurwitz-Kontsevich model \cite{hk}. Thus, despite written in the abelian approximation, matrix integral (\ref{ApproxHurwitz}) captures correctly a crucially important feature of the Hurwitz-Kontsevich model and deserves further study.

\subsection{Corrections}

Due to non-commutativity of operators ${\hat A}^i_j$, there are non-vanishing corrections at the left hand side of (\ref{GoodIdent}). We introduce them in the form of additional potential

$$ V\big(\phi, t \big) = \alpha(t) + \alpha_i(t) \ \tr \phi^i + \alpha_{ij}(t) \ \tr \phi^i \ \tr \phi^j + \alpha_{ijk}(t) \ \tr \phi^i \ \tr \phi^j \ \tr \phi^k + \ldots $$
\smallskip\\
which appears at the left hand side in

\begin{align}
\int d \phi \ \exp\left( - \dfrac{1}{2t} \tr \phi^2 + V\big(\phi, t \big) + \tr \phi {\hat A} \right) \ = \ \exp\left( \dfrac{t}{2} \ \tr {\hat A}^2 \right)
\label{GoodIdent2}
\end{align}
\smallskip\\
Potential $V\big(\phi, t \big)$ should possess a series expansion in non-negative powers of $t$

$$ \alpha_{i_1 \ldots i_m}(t) = \sum\limits_{k = 0}^{\infty} \alpha^{(k)}_{i_1 \ldots i_m} t^k $$
\smallskip\\
because subleading contributions (like those at the right hand side of (\ref{A4dev})) are always nested commutators of operators $A$ and, therefore, always have \emph{lower} $A$-degree than the leading contribution. To cancel these terms, potential $V\big(\phi, t \big)$ must contain \emph{higher} powers of $t$. The simplest terms in $V$ can be found by direct calculations, similar to (\ref{A4dev}). A parametrization, which is more convenient for these direct calculations, is

$$ e^V = \beta(t) + \beta_i(t) \ \tr \phi^i + \beta_{ij}(t) \
\tr \phi^i \ \tr \phi^j + \beta_{ijk}(t) \ \tr \phi^i \ \tr \phi^j \ \tr \phi^k + \ldots $$
\smallskip\\
where $\beta_{i_1, \ldots, i_m}$ are another parameters, which can be easily expressed through $\alpha_{i_1, \ldots, i_m}$ and vice versa. Computer experiments show, that $ \beta_{i_1, \ldots, i_m} = 0$, if $i_1 + \ldots + i_m$ is odd. For even $i_1 + \ldots + i_m$, we have
\[
\begin{array}{rl}
\\
\beta(t) = & 1 - \dfrac{t}{24} n(n^2 - 1) + \dfrac{t^2}{1152} n^2(n^2 - 1)^2 - \dfrac{t^3}{82944} n^3(n^2 - 1)^3 + O(t^4) \\
\\
\end{array}
\]
\[
\begin{array}{rl}
\\
\beta_2(t) = & \dfrac{1}{24} n - \dfrac{t}{576} n^2(n^2 - 1) + \dfrac{t^2}{27648} n^3(n^2 - 1)^2 + O(t^3) \\
\\
\beta_{11}(t) = & - \dfrac{1}{24} + \dfrac{t}{576} n(n^2 - 1) - \dfrac{t^2}{27648} n^2(n^2 - 1)^2 + O(t^3) \\
\\
\end{array}
\]
\[
\begin{array}{rl}
\\
\beta_4(t) = & -\dfrac{1}{2880} n + \dfrac{t}{69120} n^2(n^2 - 1) + O(t^2) \\
\\
\beta_{31}(t) = & \dfrac{1}{720} - \dfrac{t}{17280} n(n^2 - 1) + O(t^2) \\
\\
\beta_{22}(t) = & \dfrac{1}{5760} (5 n^2 - 6) - \dfrac{t}{138240} n(n^2 - 1) (5 n^2 - 6) + O(t^2) \\
\\
\beta_{211}(t) = & -\dfrac{1}{576} n + \dfrac{t}{13824} n^2(n^2 - 1) + O(t^2) \\
\\
\beta_{1111}(t) = & \dfrac{1}{1152} - \dfrac{t}{27648} n(n^2 - 1) + O(t^2) \\
\\
\end{array}
\]

\[
\begin{array}{lllclll}
\\
\beta_6(t) & = & \dfrac{1}{181440} n + O(t) & \ \ & \beta_{222}(t) & = & \dfrac{5n^3 - 18n}{414720}  + O(t) \\
\\
\beta_{51}(t) & = & -\dfrac{1}{30240} + O(t) & \ \ & \beta_{3111}(t) & = & -\dfrac{1}{17280} + O(t) \\
\\
\beta_{42}(t) & = & \dfrac{40 - 7n^2}{483840} + O(t) & \ \ & \beta_{2211}(t) & = & \dfrac{6 - 5 n^2}{138240} + O(t)\\
\\
\beta_{33}(t) & = & -\dfrac{1}{18144} + O(t) & \ \ & \beta_{21111}(t) & = & \dfrac{1}{27648} n + O(t)\\
\\
\beta_{411}(t) & = & \dfrac{1}{69120} n + O(t) & \ \ & \beta_{111111}(t) & = & -\dfrac{1}{82944} + O(t)\\
\\
\beta_{321}(t) & = & \dfrac{1}{17280} n + O(t) & \ \ & & \\
\\
\end{array}
\]
These terms cancel all subleading contributions up to $t^6$, i.e. they make eq. (\ref{GoodIdent2}) valid up to the order $t^6$.

\subsection{Exact matrix integral}

Above results, obtained by direct computer calculations, reveal a nice structure. To see this structure, notice that the lowest term $\beta(t)$ seems to exponentiate

$$ \beta(t) = 1 - \dfrac{t}{24} n(n^2 - 1) + \dfrac{t^2}{1152} n^2(n^2 - 1)^2 - \dfrac{t^3}{82944} n^3(n^2 - 1)^3 + \ldots \ \mathop{=}^{?} \ e^{-n(n^2-1)t/24} $$
\smallskip\\
as well as the next two terms:

$$ \beta_2(t) = \dfrac{1}{24} n \left( 1 - \dfrac{t}{24} n(n^2 - 1) + \dfrac{t^2}{1152} n^2(n^2 - 1)^2 + \ldots \right) \mathop{=}^{?} \dfrac{1}{24} n e^{-n(n^2-1)t/24} $$

$$ \beta_{11}(t) = - \dfrac{1}{24} \left( 1 - \dfrac{t}{24} n(n^2 - 1) + \dfrac{t^2}{1152} n^2(n^2 - 1)^2 + \ldots \right) \mathop{=}^{?} - \dfrac{1}{24} e^{-n(n^2-1)t/24} $$
\smallskip\\
This is clearly a hint: coefficients $\alpha$ are simpler, than coefficients $\beta$. Indeed, the simplification happens if we take a logarithm of the above series:

$$ V\big( \phi, t \big) = \log \left( \beta(t) + \beta_i(t) \  \tr \phi^i + \beta_{ij}(t) \ \tr \phi^i \ \tr \phi^j + \ldots \right) = $$

$$ = - \dfrac{t}{24} n(n^2 - 1) + \dfrac{1}{24} \tr \phi^2 \tr \phi^0 - \dfrac{1}{24} \tr \phi^1 \tr \phi^1 - \dfrac{1}{2880} \tr \phi^4 \tr \phi^0 + \dfrac{1}{720} \tr \phi^3 \tr \phi^1 - \dfrac{1}{960} \tr \phi^2 \tr \phi^2 + \emph{}$$ $$\emph{} + \dfrac{1}{181440} \tr \phi^6 \tr \phi^0 - \dfrac{1}{30240} \tr \phi^5 \tr \phi^1 + \dfrac{1}{12096} \tr \phi^4 \tr \phi^2 - \dfrac{1}{18144} \tr \phi^3 \tr \phi^3 + \mbox{ higher order terms } $$
\smallskip\\
with an obvious notation $\tr \phi^0 = n$.
Somewhat mysteriously, the first item is the central term of
the Virasoro algebra with $c = t/2$.
As one can see, the potential is simple --
much simpler than one could have expected.
It does not contain terms with more than two traces
(at least up to order $\phi^6$).
\smallskip\\

We \textbf{conjecture}, that higher order terms have the same structure:

$$ V\big( \phi, t \big) = - \dfrac{t}{24} n(n^2 - 1) + \sum\limits_{(i, j) > (0,0)}^{\infty} \alpha_{ij} \ \tr \phi^i \ \tr \phi^j $$
\smallskip\\
where the sum is taken over all non-negative $i, j$ except $0,0$. We have
\[
\begin{array}{cccccccc}
\alpha_{20} = \alpha_{02} = \dfrac{1}{48} & &\alpha_{11} = - \dfrac{1}{24} & \\
\\
\alpha_{40} = \alpha_{04} = -\dfrac{1}{5760} && \alpha_{31} = \alpha_{13} = \dfrac{1}{1440}
&& \alpha_{22} = - \dfrac{1}{960} &  \\
\\
\alpha_{60} = \alpha_{06} = \dfrac{1}{362880} & & \alpha_{51} = \alpha_{15} = - \dfrac{1}{60480}
 && \alpha_{42} = \alpha_{24} = \dfrac{1}{24192} && \alpha_{33} = - \dfrac{1}{18144}
\\ \\
\end{array}
\]
while $\alpha_{i,j}$ with odd $i + j$ vanish. Looking at these numbers, it is easy to recognize that

$$ \alpha_{ij} = \dfrac{(-1)^{ j }}{2(i + j)} \dfrac{B_{i + j}}{i! j!},
\ \ \ \ \ i+j = {\rm even \ positive}, $$
\\
where $B_{2n}$ are Bernoulli numbers:
$$B_2=\frac{1}{6},\ B_{4} = -\frac{1}{30},\ B_{6} = \frac{1}{42},
\ B_{8} = -\frac{1}{30},\ B_{10} = \frac{5}{66},
\ B_{12} = -\frac{691}{2730}, \ldots, $$
generated by $$\sum_{k=2}^\infty \frac{B_{k} z^k}{k!} = \frac{z}{e^z - 1} - 1 + \dfrac{z}{2}
\ \ \ \ {\rm or} \ \ \ \
\sum_{k=2}^\infty \frac{B_kz^k}{k!} = \frac{z}{2}\coth\left(\frac{z}{2}\right) - 1$$
Thus our conjecture is that
\begin{align}
\int d \phi \ \exp\left( - \dfrac{1}{2t} \tr \phi^2 - \dfrac{t}{24} n(n^2 - 1)
+ \sum\limits_{\stackrel{i,j = 0}{i+j\geq 2}}^\infty \dfrac{(-1)^{ j }}{2(i + j)} \dfrac{B_{i + j}}{i! j!} \
\tr \phi^i \ \tr \phi^j + \tr \phi {\hat A} \right) \ = \ \exp\left( \dfrac{t}{2} \ \tr {\hat A}^2 \right)
\label{ExactGoodIdent}
\end{align}
\smallskip\\
and exact Hurwitz-Kontsevich matrix integral is
\begin{equation}
\addtolength{\fboxsep}{5pt}
\!\boxed{
\begin{gathered}
Z_{HK} = \int\limits_{n \times n} d \phi \
\exp\left( - \dfrac{1}{2t} \tr \phi^2 - \dfrac{t}{24} n(n^2 - 1)
+ \sum\limits_{\stackrel{i,j = 0}{i+j\geq 2}}^\infty
\dfrac{(-1)^{ j }}{2(i + j)}
\dfrac{B_{i + j}}{i! j!} \ \tr \phi^i \ \tr \phi^j +
\tr \big( e^{\phi - nt/2} \psi \big) \right)
\end{gathered}
}\label{ExactHurwitz}
\end{equation}
\smallskip\\
As one can see, the Gaussian part $ - \tr \phi^2 / (2t)$
is indeed dominating in the small-$t$ limit,
because the Bernoulli part of potential is of order $t^0$
and its constant part is of order $t$.
To emphasize the $t^0$-nature of Bernoulli series
as a "quasiclassical correction" one can
simply move it to the integration measure, by summing up
the series:

$$
\sum\limits_{\stackrel{i,j = 0}{i+j\geq 2}}^\infty
\dfrac{(-1)^{ j }}{2(i + j)} \dfrac{B_{i + j}}{i! j!} \ \tr \phi^i \ \tr \phi^j
= \sum\limits_{m = 2}^{\infty} \dfrac{B_m}{2m \cdot m!} \tr
\big( \phi \otimes I - I \otimes \phi \big)^m =
\dfrac{1}{2} \tr \log \left(
\dfrac{ \sinh \left(\frac{\phi\otimes I - I\otimes\phi}{2}\right)}
{ \left(\frac{\phi\otimes I - I\otimes\phi}{2}\right)} \right)
$$
\smallskip\\
and
$$
\exp \left( \sum\limits_{\stackrel{i,j = 0}{i+j\geq 2}}^\infty
\dfrac{(-1)^{ j }}{2(i + j)} \dfrac{B_{i + j}}{i! j!} \ \tr \phi^i \ \tr \phi^j \right)
= {\det} \left(
\dfrac{ \sinh \left(\frac{\phi\otimes I - I\otimes\phi}{2}\right)}
{ \left(\frac{\phi\otimes I - I\otimes\phi}{2}\right)} \right)^{1/2}
$$
\smallskip\\
so that
\begin{equation}
\addtolength{\fboxsep}{5pt}
\boxed{
\begin{gathered}
\  Z_{HK} = e^{- \dfrac{t}{24} n(n^2 - 1)}
\int\limits_{n \times n}
\sqrt{{\det} \left(
\dfrac{ \sinh \left(\frac{\phi\otimes I - I\otimes\phi}{2}\right)}
{ \left(\frac{\phi\otimes I - I\otimes\phi}{2}\right)} \right)}\ d \phi \
\exp\left( - \dfrac{1}{2t} \tr \phi^2  +
\tr \big( e^{\phi - nt/2} \psi \big) \right) \
\end{gathered}
}\label{ExactHurwitzdet}
\end{equation}
\smallskip\\
Note that, as usual for GKM \cite{GKM,UFN3}, this integral does not depend on $n$
if expressed as a function of $p_k = \tr \psi^k$. \linebreak

It may be convenient to shift $\phi \mapsto \phi + nt/2$.
The determinantal part of the measure is obviously invariant under this shift, so we obtain
$$Z_{HK} = \int\limits_{n \times n}
\sqrt{{\det} \left(
\dfrac{ \sinh \left(\frac{\phi\otimes I - I\otimes\phi}{2}\right)}
{ \left(\frac{\phi\otimes I - I\otimes\phi}{2}\right)} \right)}\ d \phi \
\exp\left( - \dfrac{1}{2t} \tr \phi^2 - \dfrac{n}{2} \tr \phi - t \dfrac{n^3}{6}
+ t \dfrac{n}{24} + \tr \big( e^{\phi} \psi \big) \right)$$
\smallskip\\
This matrix model is, of course, of eigenvalue type.
When expressed through eigenvalues $\lambda_i$ of matrix $\phi$, the determinant part of the measure takes form

$$ {\det} \left(
\dfrac{ \sinh \left(\frac{\phi\otimes I - I\otimes\phi}{2}\right)}
{ \left(\frac{\phi\otimes I - I\otimes\phi}{2}\right)} \right)^{1/2} = \prod\limits_{i < j} \dfrac{e^{(\lambda_i - \lambda_j)/2} - e^{(\lambda_j - \lambda_i)/2}}{\lambda_i - \lambda_j} = \exp \left( \dfrac{1 - n}{2} \sum\limits_{i = 1}^{n} \lambda_i \right) \prod\limits_{i < j} \dfrac{e^{\lambda_i} - e^{\lambda_j}}{\lambda_i - \lambda_j} $$
\smallskip\\
and the angular integration can be done with the help of the Itzykson-Zuber formula \cite{Unitary}:

$$ \int [dU] \exp \Big( \tr \big( e^{\phi} U \psi U^{-1} \big) \Big) = \dfrac{ \det_{ab} \exp\left( e^{ \lambda_a } \omega_b \right) }{ \prod_{i < j} \left( e^{\lambda_i} - e^{\lambda_j} \right) \prod_{i < j} \left( \omega_i - \omega_j \right)  } $$
\smallskip\\
where $\omega_i$ are eigenvalues of matrix $\psi$. Under the integral sign, it is possible to substitute

$$ \det_{ab} \exp\left( e^{ \lambda_a } \omega_b \right) \mapsto \exp\left( \sum\limits_{i = 1}^{n} e^{ \lambda_i } \omega_i \right) $$
\smallskip\\
In this way we represent $Z_{HK}$ as $n$-fold integral over eigenvalues:

\begin{align}
\addtolength{\fboxsep}{5pt}
\!\!\! \boxed{
\begin{gathered} \!\!
Z_{HK} = \int d^n \lambda \ 
\prod_{i < j}\dfrac{\lambda_i - \lambda_j  }{ \omega_i - \omega_j }
%
\ \exp \left( - \dfrac{1}{2t} \sum\limits_{i = 1}^{n}
\lambda_i^2 - \dfrac{2n - 1}{2} \sum\limits_{i = 1}^{n} \lambda_i
- t \dfrac{n^3}{6} + t \dfrac{n}{24} + \sum\limits_{i = 1}^{n} e^{\lambda_i} \omega_i \right)\!\!
\end{gathered}
}\label{EigenHurwitz}
\end{align}
\smallskip\\
Eq.(\ref{ExactHurwitzdet}) and its eigenvalue representation (\ref{EigenHurwitz}) is
the corrected form of the naive matrix model (\ref{ApproxHurwitz}).
This is an inspiring formula, with many ingredients parallel to
intriguing observations about $Z_{HK}$ in \cite{hk},
but further discussion remains beyond the scope of the present paper.
Instead we describe in the next subsection an alternative representation
of $Z_{HK}$ in the form of a {\it discrete} matrix model \cite{UFN3}.

\subsection{Character expansion }

On the space of all polynomials in variables $p_k$,
the Hurwitz operator $\hat W_{0}$ acts in a sophisticated way,
by mixing different monomials.
However, under this sophisticated action certain polynomials --
naturally called eigenfunctions of $\hat W_{0}$ --
map into a multiple of themselves.
Actually, eigenfunctions $\chi_R$ are \emph{characters}
of irreducible representations of $GL(n)$, labeled by their signatures
$$ R = \big( \lambda_1, \lambda_2, \ldots \lambda_m \big),
\ \ \ \mbox{ with } \lambda_1 \geq \lambda_2 \geq
\ldots \geq \lambda_m $$
also known as partitions or Young diagrams.
In this section, we will use two classical formulas
from the theory of characters.

The first formula defines the eigenvalues of
the operator $\hat W_{0}$:
\begin{align}
 \hat W_{0} \chi_R = C_R \chi_R, \ \ \ \ \ \ C_R = \sum\limits_{i = 1}^{m} \lambda_i \big( \lambda_i - 2i + 1 \big)
\label{Eigenv}
\end{align}
We give a proof of this eigenvalue formula in the Appendix in s.4.5.
The first several eigenfunctions are
\[
\begin{array}{ll}
\\
\chi_{\varnothing} = 1, & \hat W_{0} \chi_{\varnothing} = 0 \\
\\
\hline
\\
\chi_{1} = p_{1}, & \hat W_{0} \chi_{1} = 0 \\
\\
\hline
\\
\chi_{2} = \dfrac{1}{2} p_{2} + \dfrac{1}{2} p_{1}^2, & \hat W_{0} \chi_{2} = + 2 \chi_{2} \\
\\
\chi_{1,1} = -\dfrac{1}{2} p_{2}+\dfrac{1}{2} p_{1}^2, & \hat W_{0} \chi_{1,1} = - 2 \chi_{1,1} \\
\\
\hline
\\
\chi_{3} = \dfrac{1}{3} p_{3}+\dfrac{1}{2} p_{2} p_{1} +
\dfrac{1}{6} p_{1}^3, & \hat W_{0} \chi_{3} = 6 \chi_{3} \\
\\
\chi_{2,1} = -\dfrac{1}{3} p_{3}+\dfrac{1}{3} p_{1}^3,
&\hat W_{0} \chi_{2,1} = 0 \\
\\
\chi_{1,1,1} = \dfrac{1}{3} p_{3}-\dfrac{1}{2} p_{2} p_{1}
+\dfrac{1}{6} p_{1}^3, & \hat W_{0} \chi_{1,1,1} = - 6 \chi_{1,1,1}\\
\\
\hline
\\
\chi_{4} = \dfrac{1}{4} p_{4}+\dfrac{1}{8} p_{2}^2
+\dfrac{1}{3} p_{1} p_{3}+\dfrac{1}{4} p_{2} p_{1}^2
+\dfrac{1}{24} p_{1}^4, & \hat W_{0} \chi_{4} = 12 \chi_{4} \\
\\
\chi_{3,1} = -\dfrac{1}{4} p_{4}-\dfrac{1}{8} p_{2}^2
+\dfrac{1}{4} p_{2} p_{1}^2+\dfrac{1}{8} p_{1}^4,
& \hat W_{0} \chi_{3,1} = 4 \chi_{3,1} \\
\\
\chi_{2,2} = \dfrac{1}{4} p_{2}^2-\dfrac{1}{3} p_{1} p_{3}
+\dfrac{1}{12} p_{1}^4, & \hat W_{0} \chi_{2,2} = 0 \\
\\
\chi_{2,1,1} = \dfrac{1}{4} p_{4}-\dfrac{1}{8} p_{2}^2
-\dfrac{1}{4} p_{2} p_{1}^2+\dfrac{1}{8} p_{1}^4,
& \hat W_{0} \chi_{2,1,1} = - 4 \chi_{2,1,1}\\
\\
\chi_{1,1,1,1} = -\dfrac{1}{4} p_{4}+\dfrac{1}{8} p_{2}^2
+\dfrac{1}{3} p_{1} p_{3}-\dfrac{1}{4} p_{2} p_{1}^2
+\dfrac{1}{24} p_{1}^4, &
\hat W_{0} \chi_{1,1,1,1} = - 12 \chi_{1,1,1,1}\\
\\
\end{array}
\]

The second formula states, that $e^{p_1}$ is decomposed
into eigenfunctions with the following coefficients:

\begin{align}
e^{p_1} = \sum\limits_{R} d_R \chi_R, \ \ \ \ \ \
d_R = \dfrac{ \prod\limits_{i < j = 1}^{m} \left( \lambda_i - \lambda_j - i + j \right) }{\prod\limits_{i = 1}^{m} \left(\lambda_i + m - i\right)!}
\label{Coefv}
\end{align}
\smallskip\\
It is well known as the \emph{hook formula} \cite{Group}. As we show in s.4.5, it is a particular case of more general Cauchy identity (\ref{Cauchy}). Up to the simple n-dependent factors $d_R$ is the dimension of
representation $R$ of $GL(n)$: 

$${\rm dim}_R = d_R \cdot \prod_{i=1}^m \frac{(n+\lambda_i-i)!}{(n-i)!}$$
\smallskip\\ 
The first several coefficients $d_R$ and corresponding dimensions ${\rm dim}_R$ are
\[
\begin{array}{ll}
\\
d_{\varnothing} = 1, & {\rm dim}_{\varnothing} = 1 \\
\\
\hline
\\
d_{1} = 1, & {\rm dim}_{1} = n \\
\\
\hline
\\
d_{2} = \dfrac{1}{2}, & {\rm dim}_{2} = \dfrac{n(n+1)}{2} \\
\\
d_{1,1} = \dfrac{1}{2}, & {\rm dim}_{1,1} = \dfrac{n(n-1)}{2} \\
\\
\hline
\\
d_{3} = \dfrac{1}{6}, & {\rm dim}_{3} = \dfrac{n(n+1)(n+2)}{6}  \\
\\
d_{2,1} = \dfrac{1}{3}, & {\rm dim}_{2,1} = \dfrac{n(n^2-1)}{3} \\
\\
d_{1,1,1} = \dfrac{1}{6}, & {\rm dim}_{1,1,1} = \dfrac{n(n-1)(n-2)}{6} \\
\\
\hline
\\
d_{4} = \dfrac{1}{24}, & {\rm dim}_{4} = \dfrac{n(n+1)(n+2)(n+3)}{24} \\
\\
d_{3,1} = \dfrac{1}{8}, & {\rm dim}_{3,1} = \dfrac{n(n+2)(n^2-1)}{8}  \\
\\
d_{2,2} = \dfrac{1}{12}, & {\rm dim}_{2,2} = \dfrac{n^2(n^2-1)}{12}\\
\\
d_{2,1,1} = \dfrac{1}{8}, & {\rm dim}_{2,1,1} = \dfrac{n(n-2)(n^2-1)}{8} \\
\\
d_{1,1,1,1} = \dfrac{1}{24}, & {\rm dim}_{1,1,1,1} = \dfrac{n(n-1)(n-2)(n-3)}{24} \\
\\
\end{array}
\]
Eqs. (\ref{Eigenv}) and (\ref{Coefv}) imply the following \emph{character expansion} for $Z_{HK}$:

\begin{equation}
\addtolength{\fboxsep}{5pt}
\boxed{
\begin{gathered}
\ \ \ Z_{HK} = \sum\limits_{R} \ e^{t C_R / 2} \ d_R \chi_R
\end{gathered}
}\label{DiscreteHurwitz}
\end{equation}
\smallskip\\
This sum over partitions, i.e. a sum over representations of $GL(n)$,
can be regarded as a \emph{discrete matrix model} \cite{UFN3}
and can be handled by the methods of
\cite{Klemm}.
Its equivalence to the matrix model (\ref{ExactHurwitzdet})
is somewhat obscure and will be discussed elsewhere.
A few first terms of the character expansion (\ref{DiscreteHurwitz}) are

\begin{align*}
Z_{HK}(p|t) \ = \ & 1 + \chi_1 + \dfrac{1}{2} e^{t} \chi_2 + \dfrac{1}{2} e^{-t} \chi_{11} + \dfrac{1}{6} e^{3t} \chi_3 + \dfrac{1}{3} \chi_{21} + \dfrac{1}{6} e^{-3t} \chi_{111} + \emph{} \\ & \dfrac{1}{24} e^{6t} \chi_4 + \dfrac{1}{8} e^{2t} \chi_{31} + \dfrac{1}{12} \chi_{22} + \dfrac{1}{8} e^{-2t} \chi_{211} + \dfrac{1}{24} e^{-6t} \chi_{1111} + \ldots
\end{align*}
\smallskip\\
or directly through the time-variables
\begin{center}
$ Z_{HK}(p|t) = $
\end{center}

\begin{center}
$1+p_{1}+\left(\dfrac{1}{4} e^{t}-\dfrac{1}{4} e^{-t}\right)p_{2}+\left(\dfrac{1}{18} e^{-3 t}-\dfrac{1}{9}+\dfrac{1}{18} e^{3 t}\right)p_{3}+\left(\dfrac{1}{96} e^{6 t}-\dfrac{1}{32} e^{2 t}+\dfrac{1}{32} e^{-2 t}-\dfrac{1}{96} e^{-6 t}\right)p_{4}+\left(\dfrac{1}{4} e^{t} + \dfrac{1}{4} e^{-t}\right)p_{1}^2 + \left(\dfrac{1}{12} e^{3 t}-\dfrac{1}{12} e^{-3 t}\right)p_{2} p_{1} + \left( \dfrac{1}{192} e^{6 t} -\dfrac{1}{64} e^{2 t}+\dfrac{1}{48}-\dfrac{1}{64} e^{-2 t}+\dfrac{1}{192} e^{-6 t}\right) p_{2}^2+\left(\dfrac{1}{72} e^{6 t} -\dfrac{1}{36}+\dfrac{1}{72} e^{-6 t}\right)p_{3} p_{1}+\left(\dfrac{1}{36} e^{3 t} + \dfrac{1}{9} + \dfrac{1}{36} e^{-3 t} \right)p_{1}^3+\left(\dfrac{1}{96} e^{6 t} + \dfrac{1}{32} e^{2 t} - \dfrac{1}{32} e^{-2 t}-\dfrac{1}{96} e^{-6 t}\right)p_{2} p_{1}^2+\left(\dfrac{1}{576} e^{6 t} + \dfrac{1}{64} e^{2 t}+\dfrac{1}{144}+\dfrac{1}{64} e^{-2 t}+\dfrac{1}{576} e^{-6 t}\right)p_{1}^4 + \ldots$
\smallskip\\
\end{center}
Note, that in this way we obtain a series in $p$-variables,
but summed in all orders of the $t$-variable.

\subsection{Appendix: $GL(n)$ characters as eigenfunctions of ${\hat W}_0$}

In this review section our starting point is the free-fermion Wick theorem
\cite{UFN3,Wick},
$$
\langle\ \prod_{i=1}^m \tilde\psi(x_i) \prod_{i=1}^m \psi(y_i)\ \rangle
= \det_{i,j}\, \langle\, \tilde \psi(x_i) \psi(y_j)\,\rangle
$$
-- a generalized form of the Fay identity, KP-equations and Shottky relations,--
which in the flat coordinates on the Riemann sphere
reduces to the elementary Cauchy identity:

$$\Delta(x) \Delta(y) \prod\limits_{i , j} \dfrac{1}{x_i - y_j}
\ =\ \det\limits_{i,j} \dfrac{1}{x_i - y_j},$$
\smallskip\\
where $\Delta(x) = \prod_{i < j} (x_i - x_j)$.
Equivalently, it can be written as

\begin{align}
\prod\limits_{i , j} \dfrac{1}{1 - x_i y_j} = \dfrac{1}{\Delta(x) \Delta(y)} \det\limits_{i,j} \dfrac{1}{1 - x_i y_j}
\label{Wick}
\end{align}
\smallskip\\
The series expansion of the right hand side has the form

$$ \dfrac{1}{\Delta(x) \Delta(y)} \det\limits_{i,j} \dfrac{1}{1 - x_i y_j} = \sum\limits_{R} \chi_R {\widetilde \chi}_R $$
\smallskip\\
where sum goes over irreducible representations
$ R = (\lambda_1 \geq \lambda_2 \geq \ldots \lambda_m) $ of $GL(n)$, and

$$ \chi_R = \chi_{\lambda_1, \ldots, \lambda_m} = \dfrac{ \det_{m \times m} \big( x^{\lambda_i + m - i}_j \big) }{\det_{m \times m} \big( x^{m - i}_j \big)}, \ \ \ \ \ \ {\widetilde \chi}_R = {\widetilde \chi}_{\lambda_1, \ldots, \lambda_m} = \dfrac{ \det_{m \times m} \big( y^{\lambda_i + m - i}_j \big) }{\det_{m \times m} \big( y^{m - i}_j \big)} $$
\smallskip\\
are the characters.
Since characters are symmetric functions in $x_i$ and $y_j$,
by inverse Miwa transform they can be expressed through
the time variables $a_k = x_1^k + \ldots + x_m^k $ and $b_k = y_1^k + \ldots + y_m^k $
(as in \cite{hk} we use the time-variables which differ from conventional ones, $t_k$,
by a $k$-factor: $p_k = kt_k$).
Expressed through $a_k$ and $b_k$, (\ref{Wick}) takes the form

\begin{align}
\exp\left( \sum\limits_{k = 1}^{\infty} \dfrac{1}{k} a_k b_k \right) = \sum\limits_{\lambda_1 \geq \lambda_2 \geq \ldots \geq \lambda_m} \chi_{\lambda_1, \ldots, \lambda_m} \big( a \big) \chi_{\lambda_1, \ldots, \lambda_m} \big( b \big) = \sum\limits_{R} \chi_{R} \big( a \big) \chi_{R} \big( b \big)
\label{Cauchy}
\end{align}
\smallskip\\
In this form, it is also known as Cauchy identity. In particular case of $a_k = \delta_{k,1}$, we recover (\ref{Coefv}):

\begin{align}
\exp\left( b_1 \right) = \sum\limits_{R} \chi_{R} \big( \delta_{k,1} \big) \chi_{R} \big( b \big)
\end{align}
\smallskip\\
where coefficients $d_R$ are given by characters

$$ d_R = \chi_{R} \big( \delta_{k,1} \big) = \dfrac{ \prod\limits_{i < j = 1}^{m} \left( \lambda_i - \lambda_j - i + j \right) }{\prod\limits_{i = 1}^{m} \left(\lambda_i + m - i\right)!} $$
\smallskip\\
The last equality requires a straightforward algebraic verification, which we do not include in this paper.

Apart from the two forms of Cauchy identity, a lot of equally explicit formulas are known for characters, see, for example, \cite{GKM,Unitary,Group}. However, Cauchy identity is quite enough for our purpose. If we put $a_k = x_1^k + \ldots + x_m^k $ and $b_k = p_k$, then (\ref{Cauchy}) turns into a "generating function" for $m$-index characters
w.r.t to the variables $x_1, \ldots, x_m$:

\begin{align}
\chi(x_1, \ldots, x_m) = \exp\left( \sum\limits_{k = 1}^{\infty} \dfrac{x_1^k + \ldots + x_m^k}{k} \ p_k \right) = \sum\limits_{\lambda_1 \geq \lambda_2 \geq \ldots \geq \lambda_m} \ \dfrac{ \det_{m \times m} \big( x^{\lambda_i + m - i}_j \big) }{\det_{m \times m} \big( x^{m - i}_j \big)} \ \chi_{\lambda_1, \ldots, \lambda_m}\big(p\big)
\label{7}
\end{align}
\smallskip\\
The simplest generating function corresponds to $m = 1$

$$ \chi(x) = \exp\left( \sum\limits_{k = 1}^{\infty} \dfrac{x^k}{k} p_k \right) = \sum\limits_{k} x^k \chi_k $$
\smallskip\\
and the next-to-simplest corresponds to $m = 2$:

$$ \chi(x, y) = \exp\left( \sum\limits_{k = 1}^{\infty} \dfrac{x^k + y^k}{k} p_k \right) = \sum\limits_{k \geq l} \left( \dfrac{ x^{k + 1} y^{l} - x^{l} y^{k + 1} }{x - y} \right) \chi_{k,l} $$
\smallskip\\
Using these generating functions as a definition of polynomials $\chi_{\lambda_1, \ldots, \lambda_m}$, we will now prove, that they are indeed eigenfunctions of the Hurwitz operator. Notice, that

$$ \dfrac{\partial}{\partial p_a} \chi(x_1, \ldots, x_m) = \sum\limits_{i = 1}^{m} \dfrac{ x_i^a }{a} \ \chi(x_1, \ldots, x_m) $$
\smallskip\\
and

$$ \dfrac{\partial}{\partial x_i} \chi(x_1, \ldots, x_m) = \sum\limits_{a = 1}^{\infty} p_a x_i^{a-1} \ \chi(x_1, \ldots, x_m) $$
\smallskip\\
Therefore
\[
\begin{array}{rlll}
{\hat W}_0 \ \chi(x_1, \ldots, x_m) & = \left( \sum\limits_{a,b = 1}^{\infty} \sum\limits_{i = 1}^{m} p_a p_b x_i^{a + b} + \sum\limits_{a,b = 1}^{\infty} \sum\limits_{i,j = 1}^{m} p_{a + b} x_i^{a} x_j^b \right) \ \chi(x_1, \ldots, x_m) \\
& \\
& = \left( \sum\limits_{a,b = 1}^{\infty} \sum\limits_{i = 1}^{m} p_a p_b x_i^{a + b} + \sum\limits_{s = 2}^{\infty} \sum\limits_{i = 1}^{m} (s - 1) p_s x_i^s + \sum\limits_{s = 2}^{\infty} \sum\limits_{i \neq j} p_{s} \dfrac{ x_j x_i^{s} - x_i x_j^{s} }{ x_i - x_j } \right) \ \chi(x_1, \ldots, x_m) \\
\\
& = \left[ \sum\limits_{i = 1}^{m} x_i^2 \dfrac{\partial^2}{\partial x_i^2} + \sum\limits_{i \neq j} \dfrac{x_i x_j}{x_i - x_j} \left( \dfrac{\partial}{\partial x_i} - \dfrac{\partial}{\partial x_j} \right) \right] \ \chi(x_1, \ldots, x_m) \\
\\
\end{array}
\]
We have just proved, that generating functions satisfy

\begin{align}
{\hat W}_0 \ \chi(x_1, \ldots, x_m) = \left[ \sum\limits_{i = 1}^{m} x_i^2 \dfrac{\partial^2}{\partial x_i^2} + \sum\limits_{i \neq j} \dfrac{x_i x_j}{x_i - x_j} \left( \dfrac{\partial}{\partial x_i} - \dfrac{\partial}{\partial x_j} \right) \right] \ \chi(x_1, \ldots, x_m)
\label{8}
\end{align}
\smallskip\\
The operator on the right hand side is a Hamiltonian of Calogero-type dynamical system. The first term in square brackets is the kinetic energy, while the second term represents interaction.
Instead of diagonalizing ${\hat W}_0$, we can do it for this Hamiltonian, which is much simpler. Let us show, that

$$ \left[ \sum\limits_{i = 1}^{m} x_i^2 \dfrac{\partial^2}{\partial x_i^2} + \sum\limits_{i \neq j} \dfrac{x_i x_j}{x_i - x_j} \left( \dfrac{\partial}{\partial x_i} - \dfrac{\partial}{\partial x_j} \right) \right] \dfrac{ \det_{m \times m} \big( x^{\lambda_i + m - i}_j \big) }{\det_{m \times m} \big( x^{m - i}_j \big)} = C_{\lambda_1, \ldots, \lambda_m} \dfrac{ \det_{m \times m} \big( x^{\lambda_i + m - i}_j \big) }{\det_{m \times m} \big( x^{m - i}_j \big)} $$
\smallskip\\
where $C_{\lambda_1, \ldots, \lambda_m}$ is a number (does not depend on $x$). This is obvious for $m = 1$

$$ x^2 \dfrac{\partial^2}{\partial x^2} x^{k} = (k^2 - k) x^{k} $$
\smallskip\\
and almost obvious for $m = 2$:

$$ \left[ x^2 \dfrac{\partial^2}{\partial x^2} + y^2 \dfrac{\partial^2}{\partial y^2} + \dfrac{2xy}{x - y} \left( \dfrac{\partial}{\partial x} - \dfrac{\partial}{\partial y} \right) \right] \left( \dfrac{ x^{k + 1} y^{l} - x^{l} y^{k + 1} }{x - y} \right) = (k^2 + l^2 - k - 3l) \left( \dfrac{ x^{k + 1} y^{l} - x^{l} y^{k + 1} }{x - y} \right) $$
\smallskip\\
For higher $m$, the proof is a straightforward algebraic exercise, which we present here in full detail. To begin with, we show that both $\Delta = \det \big( x^{m - i}_j \big)$ and $\widetilde{\Delta} = \det \big( x^{\lambda_i + m - i}_j \big)$ are eigenfunctions of kinetic energy. Indeed, making use of explicit formulas for both determinants

$$\widetilde{\Delta} = \det_{m \times m} \big( x^{\lambda_i + m - i}_j \big) = \sum\limits_{\sigma \in S_{m}} (-1)^{|\sigma|} x_{1}^{\lambda_{\sigma_1} + m - \sigma_1} \ldots x_{m}^{\lambda_{\sigma_m} + m - \sigma_m} $$
\smallskip\\
and
$$\Delta = \det_{m \times m} \big( x^{m - i}_j \big) = \sum\limits_{\sigma \in S_{m}} (-1)^{|\sigma|} x_{1}^{m - \sigma_1} \ldots x_{m}^{m - \sigma_m} $$
\smallskip\\
we have

$$\sum\limits_{i = 1}^{m} x_i^2 \dfrac{\partial^2 \widetilde{\Delta}}{\partial x_i^2} = \sum\limits_{\sigma \in S_{m}} (-1)^{|\sigma|} \sum\limits_{i = 1}^{m} \big( \lambda_{\sigma_i} + m - \sigma_i \big) \big( \lambda_{\sigma_i} + m - \sigma_i - 1 \big) \ x_{1}^{\lambda_{\sigma_1} + m - \sigma_1} \ldots x_{m}^{\lambda_{\sigma_m} + m - \sigma_m} $$
\smallskip\\
The sum over $i$, which appears in the right hand side, is invariant under permutations and factors out:

$$\sum\limits_{i = 1}^{m} x_i^2 \dfrac{\partial^2 \widetilde{\Delta}}{\partial x_i^2} = \sum\limits_{i = 1}^{m} \big( \lambda_i + m - i \big) \big( \lambda_i + m - i - 1 \big) \ \widetilde{\Delta} $$
\smallskip\\
Similarly

$$\sum\limits_{i = 1}^{m} x_i^2 \dfrac{\partial^2 \Delta}{\partial x_i^2} = \sum\limits_{\sigma \in S_{m}} (-1)^{|\sigma|} \sum\limits_{i = 1}^{m} \big( m - \sigma_i \big) \big( m - \sigma_i - 1 \big) x_{1}^{m - \sigma_1} \ldots x_{m}^{m - \sigma_m} = \sum\limits_{i = 1}^{m} \big( m - i \big) \big( m - i - 1 \big) \ \Delta $$
\smallskip\\
So $\Delta$ and $\widetilde{\Delta}$ are eigenfunctions of kinetic energy. By straightforward differentiation, we obtain

$$ \sum\limits_{i = 1}^{m} x_i^2 \dfrac{\partial^2 }{\partial x_i^2} \left( \dfrac{\widetilde{\Delta}}{\Delta} \right) = \dfrac{1}{\Delta} \sum\limits_{i = 1}^{m} x_i^2 \dfrac{\partial^2 \widetilde{\Delta}}{\partial x_i^2} - \dfrac{\widetilde{\Delta}}{\Delta^2} \sum\limits_{i = 1}^{m} x_i^2 \dfrac{\partial^2 \Delta}{\partial x_i^2} - \dfrac{2}{\Delta^2} \sum\limits_{i = 1}^{m} x_i^2 \dfrac{\partial \widetilde{\Delta}}{\partial x_i} \dfrac{\partial \Delta}{\partial x_i} + \dfrac{2}{\Delta^3} \sum\limits_{i = 1}^{m} x_i^2 \widetilde{\Delta} \dfrac{\partial \Delta}{\partial x_i} \dfrac{\partial \Delta}{\partial x_i} = $$
\smallskip\\

$$ = \sum\limits_{i = 1}^{m} \left( \lambda_i + 2m - 2i - 1 \right) \ \dfrac{\widetilde{\Delta}}{\Delta} - \dfrac{2}{\Delta^2} \sum\limits_{i = 1}^{m} x_i^2 \dfrac{\partial \widetilde{\Delta}}{\partial x_i} \dfrac{\partial \Delta}{\partial x_i} + \dfrac{2}{\Delta^3} \sum\limits_{i = 1}^{m} x_i^2 \widetilde{\Delta} \dfrac{\partial \Delta}{\partial x_i} \dfrac{\partial \Delta}{\partial x_i}$$
\smallskip\\
Including interactions and considering the full Hamiltonian, we obtain
\[
\begin{array}{cc}
\\
\left[ \sum\limits_{i = 1}^{m} x_i^2 \dfrac{\partial^2}{\partial x_i^2} + \sum\limits_{i \neq j} \dfrac{x_i x_j}{x_i - x_j} \left( \dfrac{\partial}{\partial x_i} - \dfrac{\partial}{\partial x_j} \right) \right] \ \dfrac{\widetilde{\Delta}}{\Delta} - \sum\limits_{i = 1}^{m} \lambda_i \big( \lambda_i + 2m - 2i - 1 \big) \ \dfrac{\widetilde{\Delta}}{\Delta} = \\ \\ = \sum\limits_{i \neq j} \dfrac{2 x_i x_j}{x_i - x_j} \left( \dfrac{1}{\Delta} \dfrac{\partial \widetilde{\Delta}}{\partial x_i} - \dfrac{\widetilde{\Delta}}{\Delta^2} \dfrac{\partial \Delta}{\partial x_i} \right) - \dfrac{2}{\Delta^2} \sum\limits_{i = 1}^{m} x_i^2 \dfrac{\partial \widetilde{\Delta}}{\partial x_i} \dfrac{\partial \Delta}{\partial x_i} + \dfrac{2}{\Delta^3} \sum\limits_{i = 1}^{m} x_i^2 \widetilde{\Delta} \dfrac{\partial \Delta}{\partial x_i} \dfrac{\partial \Delta}{\partial x_i} \\ \\
\end{array}
\]
The sum over $j$ is easily evaluated

$$ \sum\limits_{j \neq i} \dfrac{2 x_i x_j}{x_i - x_j} = x_i \left( - 2 m + 2 + \dfrac{2 x_i}{\Delta} \dfrac{\partial \Delta}{ \partial x_i} \right) $$
\smallskip\\
and, after cancelation of terms, we obtain
\[
\begin{array}{cc}
\left[ \sum\limits_{i = 1}^{m} x_i^2 \dfrac{\partial^2}{\partial x_i^2} + \sum\limits_{i \neq j} \dfrac{x_i x_j}{x_i - x_j} \left( \dfrac{\partial}{\partial x_i} - \dfrac{\partial}{\partial x_j} \right) \right] \ \dfrac{\widetilde{\Delta}}{\Delta} - \sum\limits_{i = 1}^{m} \lambda_i \big( \lambda_i + 2m - 2i - 1 \big) \ \dfrac{\widetilde{\Delta}}{\Delta} = \\ \\ = \sum\limits_{i = 1}^{m} x_i \left( - 2 m + 2 + \dfrac{2 x_i}{\Delta} \dfrac{\partial \Delta}{ \partial x_i} \right) \left( \dfrac{1}{\Delta} \dfrac{\partial \widetilde{\Delta}}{\partial x_i} - \dfrac{\widetilde{\Delta}}{\Delta^2} \dfrac{\partial \Delta}{\partial x_i} \right) - \dfrac{2}{\Delta^2} \sum\limits_{i = 1}^{m} x_i^2 \dfrac{\partial \widetilde{\Delta}}{\partial x_i} \dfrac{\partial \Delta}{\partial x_i} + \dfrac{2}{\Delta^3} \sum\limits_{i = 1}^{m} x_i^2 \widetilde{\Delta} \dfrac{\partial \Delta}{\partial x_i} \dfrac{\partial \Delta}{\partial x_i} = \\ \\
= \sum\limits_{i = 1}^{m} x_i \left( - 2 m + 2 \right) \left( \dfrac{1}{\Delta} \dfrac{\partial \widetilde{\Delta}}{\partial x_i} - \dfrac{\widetilde{\Delta}}{\Delta^2} \dfrac{\partial \Delta}{\partial x_i} \right) = \left( - 2 m + 2 \right) \left( \deg \widetilde{\Delta} - \deg \Delta \right) \ \dfrac{\widetilde{\Delta}}{\Delta} \\
\end{array}
\]
\smallskip\\
By definition, $\deg \widetilde{\Delta} = \sum\limits_{i = 1}^{m} (\lambda_i + i - 1)$ and $\deg \Delta = \sum\limits_{i = 1}^{m} (i - 1)$. So, we have proved the identity

\begin{align}
 \left[ \sum\limits_{i = 1}^{m} x_i^2 \dfrac{\partial^2}{\partial x_i^2} + \sum\limits_{i \neq j} \dfrac{x_i x_j}{x_i - x_j} \left( \dfrac{\partial}{\partial x_i} - \dfrac{\partial}{\partial x_j} \right) \right] \ \dfrac{\widetilde{\Delta}}{\Delta} = \sum\limits_{i = 1}^{m} \lambda_i \big( \lambda_i - 2i + 1 \big) \ \dfrac{\widetilde{\Delta}}{\Delta}
\label{9}
\end{align}
\smallskip\\
thus diagonalizing this Hamiltonian. At the same time, via Cauchy identity, we have proved the dual result

\begin{align}
\sum_{a,b = 1}^{\infty} \left((a+b)p_ap_b\frac{\p}{\p p_{a+b}} + abp_{a+b}\frac{\p^2}{\p p_a\p p_b}\right) \chi_{\lambda_1, \ldots, \lambda_m}\big(p \big) = \sum\limits_{i = 1}^{m} \lambda_i \big( \lambda_i - 2i + 1 \big) \ \chi_{\lambda_1, \ldots, \lambda_m}\big(p \big)
\label{10}
\end{align}
\smallskip\\
thus diagonalizing the $W_0^{(3)}$ operator with quadratic eigenvalues

\begin{align}
C_{\lambda_1, \ldots, \lambda_m} = \sum\limits_{i = 1}^{m} \lambda_i \big( \lambda_i - 2i + 1 \big)
\label{11}
\end{align}
\smallskip\\
Above calculation is somewhat tedious, but important:
it shows, that the $\hat W_0^{(3)}$ "cut-and-join" operator,
defining the generating function of Hurwitz numbers,
can be rewritten as a Calogero-type Hamiltonian of multi-particle system in 1d
with coordinates $x_i$ and pairwise interactions between particles.

\section{Conclusion}

In this paper we considered three formulas, relating the two {\it a priori}
unrelated kinds of objects: partition functions of the matrix models and generators
of the $W$-algebra.

Eq.(\ref{Gauss}) provides a one more representation for Hermitian
matrix model -- the main personage of all matrix-model studies.

Eq.(\ref{3}) describes a mush less investigated version of the same model,
with extra background field $\psi$. The two $W$-operators in the two formulas
(\ref{3}) and (\ref{Gauss}) act on the two conjugate sets of time-variables
in Hermitian model. It would be interesting to extend these expressions
to non-Gaussian, say, DV phases of the theory.

Finally, eqs.(\ref{4}) and (\ref{ExactHurwitzdet}) provide an
inspiring matrix-model representation for the Hurwitz-Kontsevich
tau-function -- a recently discovered new link between combinatorics
and integrability theory.
It would be very interesting to explicitly describe the intriguingly
sophisticated Virasoro constraints for this partition function as
the Ward identities for this matrix model.

\section*{Acknowledgements}

We are indebted to A.Mironov for important comments. Our work is partly supported by Russian Federal Nuclear Energy Agency
and the Russian President's Grant of Support for the Scientific Schools NSh-3035.2008.2, by RFBR grant 07-02-00645,
by the joint grants 09-01-92440-CE, 09-02-91005-ANF, 09-02-93105-CNRS
and by the NWO project 047.011.2004.026.
The work of Sh.Shakirov is also supported in part by
the Moebius Contest Foundation for Young Scientists and by the Dynasty Foundation.

\end{document}